\documentclass{egpubl}
\usepackage{eurovis2023}

\EuroVisShort  

\usepackage[T1]{fontenc}
\usepackage{dfadobe}  

\usepackage{cite}  
\BibtexOrBiblatex
\electronicVersion
\PrintedOrElectronic
\ifpdf \usepackage[pdftex]{graphicx} \pdfcompresslevel=9
\else \usepackage[dvips]{graphicx} \fi

\usepackage{egweblnk}
\usepackage{xcolor}
\usepackage{amsmath}
\usepackage{colortbl}
\usepackage{subcaption}
\usepackage{algorithm}
\usepackage{algpseudocode}
\usepackage{amsfonts}

\newcommand{\FA}[1]{\colorbox{blue!15}{#1}}
\newcommand{\SM}[1]{\colorbox{gray!15}{#1}}
\newcommand{\FAC}[1]{\cellcolor{blue!15}{#1}}
\newcommand{\FACS}[1]{\cellcolor{blue!35}{#1}}
\newcommand{\SMC}[1]{\cellcolor{gray!15}{#1}}
\newcommand{\SMCS}[1]{\cellcolor{gray!35}{#1}}

\newcommand{\prob}{cluttering }

\title[Cluttering Edges Heuristic]{Identifying Cluttering Edges in Near-Planar Graphs}

\author[S. van Wageningen \& T. Mchedlidze \& A. Telea]
{\parbox{\textwidth}{\centering S. van Wageningen$^{1}$\orcid{0000-0002-0346-5597} and T. Mchedlidze$^{1}$\orcid{0000-0001-6249-3419} and  A. Telea$^{1}$\orcid{0000-0003-0750-0502} 
    }\\
{\parbox{\textwidth}{\centering $^1$Utrecht University, The Netherlands\\
       }
}
}

\begin{document}


\teaser{
\vspace{-1cm}
\includegraphics[width=\linewidth,height=0.2\linewidth]{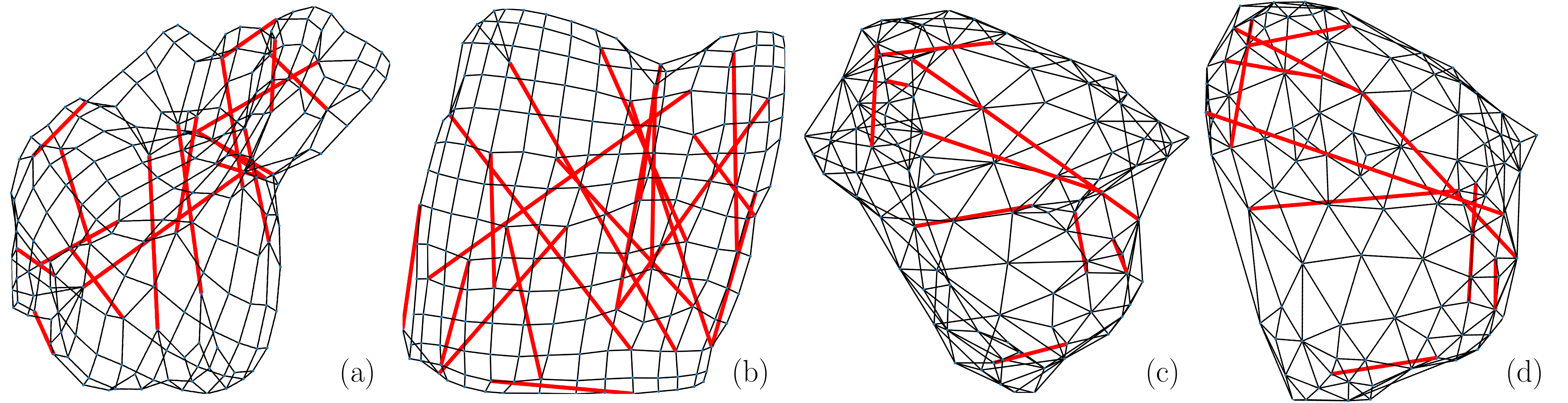}
{\phantomsubcaption\label{fig:teaser_grid1}\phantomsubcaption\label{fig:teaser_grid2}\phantomsubcaption\label{fig:teaser_del1}\phantomsubcaption\label{fig:teaser_del2}}
\centering
\caption{Graphs laid out by ForceAtlas2 with augmenting edges colored red. (a) augmented grid (\texttt{redraw}), (b) augmented grid weighted by the \texttt{heuristic} ($H_\textrm{min}$), (c) triangulations (\texttt{redraw}), (d) triangulations weighted by the \texttt{heuristic} ($H_\textrm{min}$)}
\label{fig:teaser}
}

\maketitle
\begin{abstract}
Planar drawings of graphs tend to be favored over non-planar drawings. Testing planarity and creating a planar layout of a planar graph can be done in linear time. However, creating readable drawings of nearly planar graphs remains a challenge. We therefore seek to answer which edges of nearly planar graphs create clutter in their drawings generated by the mainstream graph drawing algorithms. We present a heuristic to identify problematic edges in nearly planar graphs and adjust their weights, in order to produce higher quality layouts with spring-based drawing algorithms. Our experiments show that the heuristic produces significantly higher quality drawings for augmented grid graphs, augmented triangulations and deep triangulations.

\ccsdesc[500]{Human-centered computing~Graph drawings}

\printccsdesc   
\end{abstract}  
\section{Introduction}

The ultimate goal when constructing a readable drawing of a graph (i.e. node-link diagram) is to avoid clutter that prevents viewer from grasping the graph's structure. One of the quality metrics that measures the clutter of a graph drawing is the number of edge crossings. It is long known that humans perform better on shortest path related tasks in drawings with fewer crossing~\cite{planar_favored}. Humans also tend to prefer such drawings~\cite{crossing_preference,crossing_preference_two}.  It is therefore natural to request that a drawing of a graph possesses no edge intersections at all, whenever possible. Such drawings, and the graphs that can be drawn in such a way, are called \emph{planar}. Fortunately, detecting whether a graph has a planar drawing~\cite{planar_testing} and constructing one in affirmative\cite{handbook_planar} can be done in linear time. 
In practice, on the other hand, graphs are rarely planar. However, they can still be sparse, or contain clear planar substructures, in other words, be \emph{nearly planar}. It is therefore desirable for such graphs to achieve \emph{nearly planar} drawings.  

There are various attempts to formalize the notion of \emph{near-planarity} and construction of nearly planar drawing. Unfortunately, all of these attempts lead to hard computational problems~\cite{planar_np,crossing_NP_hard_near_planar,RAC_NP,one_planar_NP_complete}. There are a few spring-based algorithms that address certain readability issues relevant to near-planarity, such as trying to keep crossing angles close to $90^\circ$~\cite{ArgyriouBS12} or improving layouts by preserving existing crossings~\cite{simonetto2011impred}. Yet, there is a lot of room for further work to generate practical working algorithms that construct readable drawings of near planar graphs.

In this paper we propose a spring-based heuristic approach to construct nearly planar drawing of graphs that contain dense planar substructures. This work is motivated by the lack of comparable approaches, and the aforementioned hardness of formal definitions of near-planarity. We conduct an experimental evaluation comparing our approach to the state-of-the-art spring-based algorithms. 

The paper is structured as follows. In Section~\ref{sec:initial_exp}, we present initial observations on how to resolve clutter in near planar drawings, and provide an initial experiment that lies in the background of our approach. In Section~\ref{sec:approach}, we present our heuristic. Section~\ref{sec:setup} describes the experimental setup, whereas Section~\ref{sec:results} presents the results of our experiments. 

\section{Related work}\label{sec:related_work}
Throughout the paper we denote by $G=(V,E,w)$ a graph, with $V$, $E$ being the sets of nodes, edges, respectively and $w:E \rightarrow \mathbb{R}$ the edge-weighting function. Here $n=|V|$ and $m=|E|$. For nodes $u$ and $v$, we denote by $e=\{u,v\}$ and $e=(u,v)$, an undirected edge and a directed edge.

\paragraph*{Theoretical approaches to near-planarity}
Since a planar drawing is the one that does not contain any crossing, the most straight-forward idea to define near-planarity is to request a drawing with as few crossings as possible~\cite{crossing_NP_hard_near_planar}; this problem is NP-hard~\cite{planar_np} even for a planar graph plus a single edge~\cite{crossing_NP_hard_near_planar}. There are a few algorithms that insert edges into planar graphs and their drawings in a crossing optimal way~\cite{RadermacherR22,GutwengerKM08,GutwengerMW05,ChimaniH16,planarization_impr}, of which have been compared experimentally~\cite{GutwengerM03}. 
 
While we know that humans perform tasks well on planar layouts~\cite{planar_favored}, it has been also  shown experimentally that the negative effects of the crossings on the task performance decreases as the edge crossing angles increase~\cite{crossing_angle_task_performance}. This led to the definition of RAC~\cite{RAC} and $\alpha$-AC drawings~\cite{GiacomoDLM11}. Deciding whether such drawings exist is NP-hard as well~\cite{RAC_NP}. If the graph is sparse it is also natural to try to limit how many crossings an edge has, as fewer crossings would impair less the perception of the edge. This idea led to definition of \emph{k-planar} and \emph{quasi-planar} drawings. However, again, recognizing whether a graph has these kind of drawings is NP-complete \cite{one_planar_NP_complete}.

\paragraph*{Heuristics}
A plethora of spring-based algorithms produce high-quality layouts without specifically targeting nearly planar graphs~\cite{handbook_spring}. Out of these approaches, we refer specifically to ForceAtlas2 (\emph{FA2})~\cite{FA2,FA2_python} and Stress Majorization (\emph{SM})~\cite{stress_majorization} which are powerful layout techniques that we use for our experiments, since they are often compared with novel techniques~\cite{deepdrawing,MLapproach}. SM solely bases the spring forces between all pairs of nodes on the length of their shortest paths, whereas FA2 considers attractive and repulsive forces to compute node spring forces. Additionally, the work on SM~\cite{stress_majorization} suggests to weight node pairs by taking into account the number of common neighbors. Here, the weight of a node-pair $u,v$ is set as: $w(u,v)= |N_u \cup N_v| - |N_u \cap N_v|$, where $N_{u}$ denotes the neighborhood of node $u$. This idea improves the performance of SM when edge lengths need to vary significantly. As will become clear in the following, this approach is relevant to ours and is therefore included as the \emph{neighborhood weighting function} in our experiments.

 Finally, relevant to near-planarity, Argyriou et al.~\cite{ArgyriouBS12} present an approach that maximizes the \emph{total resolution}, which is the minimum of the \emph{angular} and \emph{crossing} resolution. While ImPrEd~\cite{simonetto2011impred} preserves the topology of the given layout and therefore its planarity. Similarly, tsNET$^\star$~\cite{Kruiger2017_tsnet} tends to preserve a layout's original structure by favoring occasional long edges and thus unfolding the layout.

\section{Preliminaries} 
\label{sec:initial_exp}
Our overarching goal is to produce a drawing of a nearly planar graph $G$ which clearly depicts its planar substructure. This statement itself hints us to
distinguish among the graph edges that contribute to its planar substructure and those that destroy it. If we were able to detect the latter edges, we could remove them, construct a planar drawing of the remaining graph (using e.g.~\cite{FraysseixPP90,Schnyder90}), and place the earlier removed edges back. There are two challenges that prevent us from taking this approach. The first one is that detecting a dense planar substructure is a hard optimization problem known as \emph{Maximum planar subgraph}~\cite{Cimikowski95}. The second is that such an approach would inevitably be based on algorithms to construct planar drawings of planar graphs, e.g.~\cite{FraysseixPP90,Schnyder90}, which are relatively hard to implement and up to date are not part of most graph drawing libraries and applications. Therefore, we chose to attack our problem using a relatively simple heuristic based on a spring-based approach.

An initial idea was to identify such \emph{planarity-destroying} edges. Once identified, we would weight them with relatively lower weights than regular edges. We  would then use a state-of-the-art spring-based approach, that takes edge-weights into account. Our hope was that the planarity-destroying edges would influence the layout less than the remaining edges and therefore the planar substructure would reveal itself in the drawing.

\begin{figure}[t!]
    \includegraphics[width=\linewidth,height=0.4\linewidth]{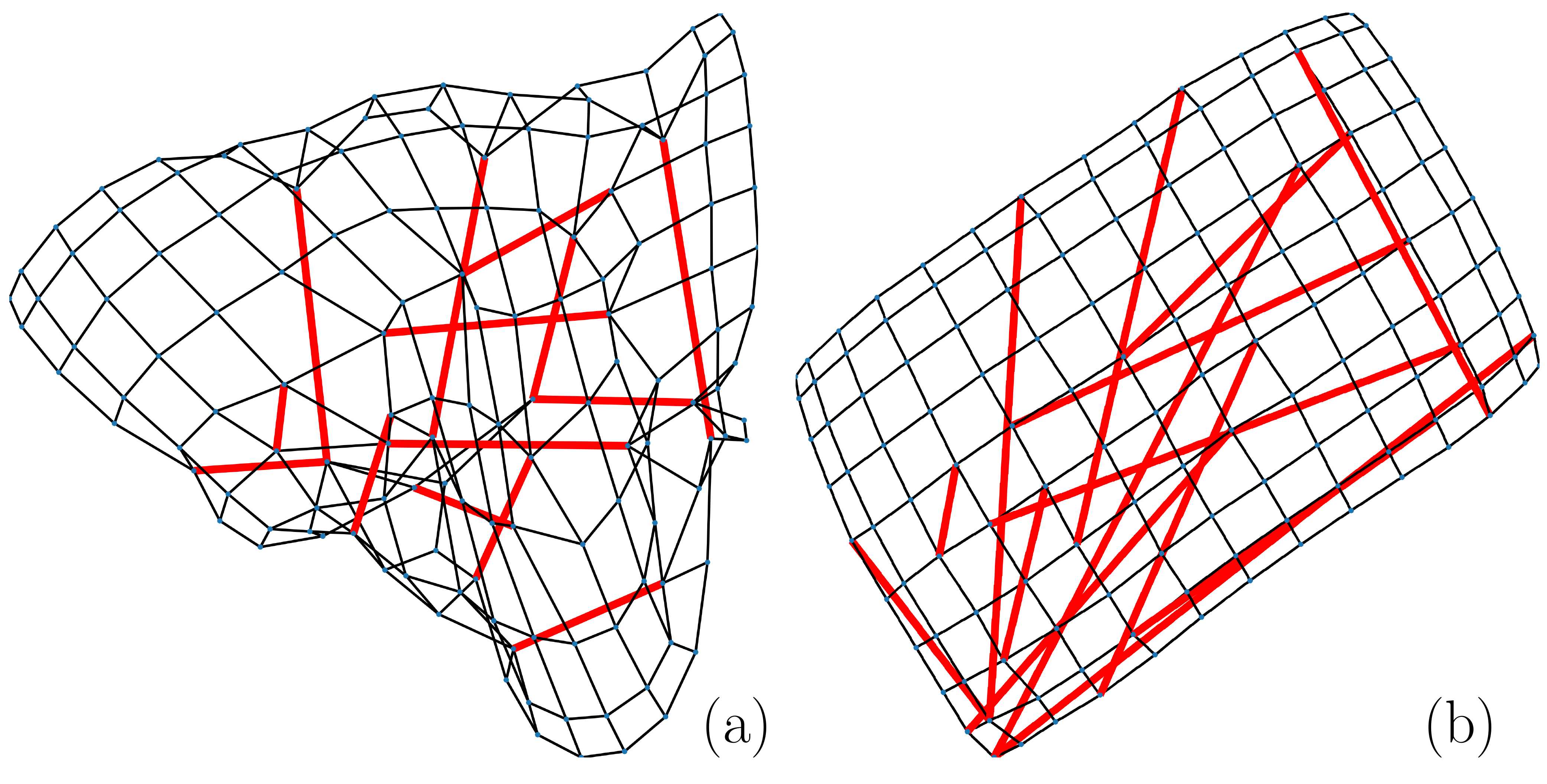}

    {\phantomsubcaption\label{fig:aug_grid_redrawn}\phantomsubcaption\label{fig:aug_grid_rweighted}}

\caption{(a) Augmented grid (b), weighted augmented grid}
\vspace{-1cm}
\label{fig:example_grid_redrawn}
\end{figure}

To test whether this idea is feasible, we perform the following initial experiment. We consider a grid $D=(V,E)$ and construct a graph $G=(V,E\bigcup E_p)$, where $E_p$ is a set of random edges on the vertex set $V$ with the property that $E \bigcap E_p = \emptyset$. We call this graph an \emph{augmented grid}. Starting with a random initial coordinate assignment, we draw an augmented grid using FA2. These layouts (see Figure~\ref{fig:aug_grid_redrawn}), appear cluttered and folded inwards, and of poor quality. We then reduce the weights of the edges in $E_p$ to $0.01$ and rerun FA2. The resulting layout 
(Figure~\ref{fig:aug_grid_rweighted}) has unfolded and has become a near perfect depiction of a grid graph. This experiment hints that knowing the planarity-destroying edges can be useful in depicting planar substructures in nearly planar graphs, using a spring-based approach and appropriate edge weighting scheme.

In our experiments with other augmented planar graph classes, we observe that not only the planarity-destroying edges create clutter in the drawing, but so do the edges that are close to the outer face.
These observations led to the following questions: Which edges of a given nearly planar graph $G$ create clutter in the drawings of $G$ generated by a spring-embedding algorithm? Using a spring-based approach, how do we weight these edges to create a drawing where a planar substructure is clearly visible?
We call such edges \emph{\prob} and address the stated questions in the following section. The Python implementation of the heuristic can be found on \href{https://github.com/simonvw95/beyond_planarity_sb}{GitHub}. 


\section{Finding and weighting \prob edges}
\label{sec:approach}
\paragraph*{Vertex-disjoint paths}
Our approach to identify \prob edges is based on the following intuition. If the end-vertices 
of an edge $e=\{u,v\}$ are connected by multiple, relatively short paths in the graph, then the edge $e$ can also be short. In the opposite case, where there are only relatively long paths between $u$ and $v$, then the edge $e$ could collapse the drawing and therefore could be a \prob one.  However, finding all paths between even a single pair of vertices will result in
exponential computations. Hence, we use as a proxy the lengths of the vertex-disjoint paths between $u$ and $v$.  

\begin{figure}[t!]
\centering
   \includegraphics[width=0.5\linewidth]{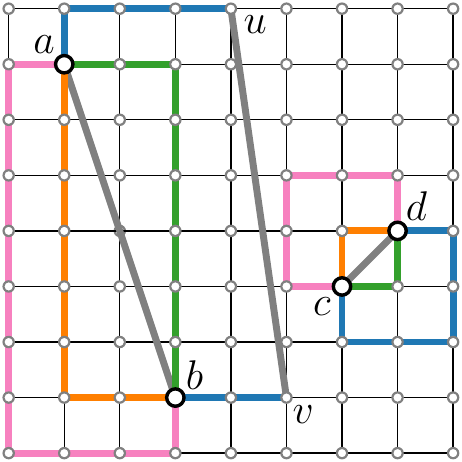}
    \caption{A grid with three gray augmenting edges $\{a,b\}$, $\{c,d\}$ and $\{u,w\}$. Vertex-disjoint paths are depicted by colors. The footprints of   $\{a,b\}$, $\{c,d\}$ are
    $f(\{c,d\}) = [2,2,6,6]$, $f(\{a,b\}) = [7,8,8,12]$. Note that a footprint of edge $\{c,d\}$, that need to be short in a grid-revealing layout, contains small values. While the footprint of the long edge $\{a,b\}$ contains mostly large values. Observe that footprint of $\{a,b\}$ uses another augmenting edge and thus can also contain smaller numbers but they are relatively fewer and larger than the numbers in the footprints of short edges.} 
    \label{fig:footprint}
\end{figure}

Vertex-disjoint paths can be found using the max-flow Edmonds-Karp algorithm \cite{edmonds_karp}, as described in~\cite{KleinbergT06}. For completeness, we summarize the idea of this reduction here. A directed graph with edge-weights of value $1$ has $k$ edge independent paths between two vertices $u$ and $v$ if and only if it has a flow of value $k$ between $u$ and $v$. To reduce the undirected problem to the directed, we substitute every undirected edge $\{u,v\}$ by two $(u,v)$ and $(v,u)$. To ensure that the paths are vertex disjoint, we substitute every node $u$ by two $u_s$ and $u_t$, add to the graph directed edge $(u_s,u_t)$ and substitute every edge $(w,u)$, incoming to $u$, by $(w,u_s)$; analogously, substitute every edge $(u,w)$, outgoing from $u$, by $(u_t,w)$. In the following, for an edge $e=\{u,v\}$, we denote by $f(e)=[\ell_1,\ell_2,\dots]$ the sequence of the lengths of the vertex-disjoint paths between $u$ and $v$ in $G\setminus e$,
 listed in non-decreasing order, and call $f(e)$ the \emph{footprint} of $e$. 
Given that the complexity of Edmond-Karp's algorithm is $O(VE^2)$, computation of footprints for all edges of $G$ takes $O(VE^3)$ time. 

\paragraph*{Outlier detection}
By analyzing the footprints of edges in augmented grids, we observe patterns that distinguish the edges that need to stay short in a grid-revealing layout and those that need to get long in such layouts. Refer to Figure~\ref{fig:footprint} for a detailed example. 
These differences in the footprints  brings us the idea of using an outlier detection algorithm to detect \prob edges.
We experiment with the \emph{Isolation Forest} technique~\cite{isolation_forest} which is designed to find anomalous data points in n high-dimensional space. Intuitively, the Isolation Forest method isolates a single specific data point by creating partitions through selecting random cutoff points along different dimensions. Once the data point is completely isolated, the number of partitions inform whether the point is anomalous or not. Few required partitions indicate an anomalous behavior. This partitioning process is applied to every data point, and subsequently repeated to get an average estimate. The time-complexity of the technique is $\mathcal{O}(s(m+p)p)$, where $s$ is the number of times the technique is repeated, and $p$ is the number of partitions needed to isolate a node. The Isolation Forest is a simple but effective technique, and can be applied to detect \emph{outlier} edges according to their footprints in linear time .

\paragraph*{Footprint normalization and \prob edge weighting}
Before however applying the Isolation Forest method to footprints of the edges, we have to make sure that they all have the same length. For this purpose, we expand or contract the footprints, depending on the user-specified number of dimensions $k$ and function $\mathcal M$, which can be either the minimum, maximum or mean function. Equation \ref{eq:fv} portrays how the footprints are expanded or contacted, given a footprint $f(e)$ of initial length $l$ and a desired length $k$. 

\begin{equation}
\label{eq:fv}
f'(e) = \begin{cases}
f(e) \oplus [{\mathcal M}(f(e))]_{k-l} & l < k \\
f(e) & l = k \\
f(e)[0:k-1] \oplus {\mathcal M}(f(e)[k:l]) & l > k \\ 
\end{cases}
\end{equation}

In our experiments we evaluate the results for all aforementioned choices of function $\mathcal M$, i.e. minimum, maximum or mean. Based on multiple experiments, we set the weight of edge $e$ to $\mathcal{M}(f(e))$. 

\section{Experimental setup}\label{sec:setup}
\begin{table}[t!]
\centering
\small
\begin{tabular}{||l||l|l|l|l|l||}
\hline
 \emph{Dataset} & $n_{min}$ & $n_{max}$ & $m_{min}$ & $m_{max}$ & \#\emph{Graphs}\\
\hline
\hline
 \textbf{Grids} & 48 & 196 & 86 & 383 & 50\\
 \textbf{Triangulations} & 26 & 100 & 69 & 297 & 50\\
 \textbf{Deep triangulations} & 25 & 98 & 86 & 367 & 50\\
 \textbf{Rome} & 18 & 100 & 25 & 141 & 100\\
\hline
\end{tabular}
\caption{Statistics of datasets used in the experiments\label{tab:dataset_description}}
\vspace{-0.5cm}
\end{table}

Let $G=(V,E)$ be a nearly planar graph. To evaluate our approach, we apply 
spring-based algorithms to the weighted graph $G_\mathcal{M}=(V,E,w)$, $w(e)=\mathcal{M}(f(e))$ and the baseline unweighted graph $G=(V,E)$. We compare the obtained drawings both qualitatively, by exploring them visually, and quantitatively, by employing several quality metrics. In the following, we discuss the datasets used for the experiment, the way the layouts are computed, and the measured quality metrics. 

\paragraph*{Datasets}\label{sec:datasets}
We used four types of graphs, as follows (see Table~\ref{tab:dataset_description}). 

\noindent\textbf{Grids:} We start with a grid with a random number 
of rows and columns ranging between $6$ and $14$.
Next, we add $0.1|V|$ edges 
between random non-adjacent nodes to destroy the grid structure, yielding the \emph{augmented grid}. Note that most, but not every, such added edge introduces edge crossings.

\smallskip
\noindent\textbf{Triangulations:} These graphs are generated by applying
Delaunay triangulation algorithm (implemented in SciPy) on random 2D point sets, to which structure-destroying edges are added similarly as to the Grids. We call these graphs \emph{augmented triangulations}. For a grid or a triangulation $G$, we denote by $\overline{G}$ its augmented version.  

\smallskip 
\noindent\textbf{Deep triangulations:} Even planar graphs can be a challenge for spring-based approaches when it comes to unraveling their planar structure, especially when the planar-layout edges need to have various lengths. This is the case for planar graphs that contain small cycles with many vertices inside them. To further test  our approach, we construct so called \emph{deep triangulations}, as follows: (1) Randomly place $0.7|V|$ of the vertices and construct their Delaunay triangulation $T$. (2) Place a random number of points $r < 0.3|V|$ in a random triangle $t \in T$. (3) Apply Delaunay triangulation of the $r$ points in $t$ to create new edges. (4) Repeat steps 2 and 3 steps until all remaining $0.3|V|$ vertices have been placed.

\smallskip

\noindent\textbf{Rome:} We also include a subset of non-planar \emph{Rome} graphs \cite{rome} into our experiment. Note that grids and (deep) triangulations contain dense planar substructures and therefore we expect our heuristic to be able to find \prob edges in these graphs. However, we also include 
Rome graphs to our experiments, to check whether our technique generalizes to this fairly popular graph benchmark~\cite{deepgd}, which however are very sparse and do not contain dense planar substructures. 

\begin{figure}[t!]
    \centering
     \begin{subfigure}{0.425\textwidth}
        \includegraphics[width=1.0\linewidth,height=0.8\linewidth]{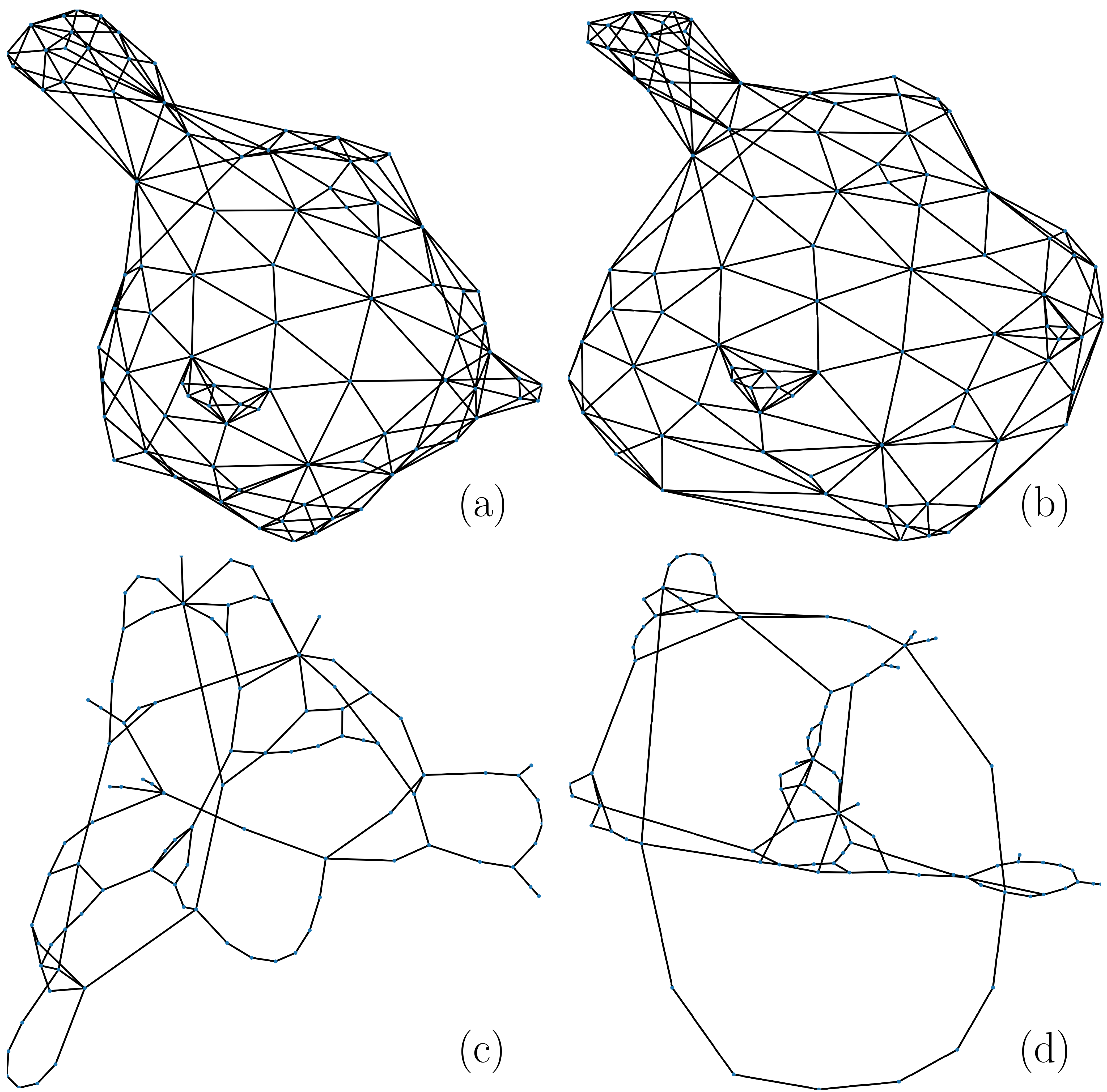}
        \phantomcaption
        \label{fig:deep_a}
    \end{subfigure}
    \begin{subfigure}{0.0\textwidth}
        \phantomcaption
        \label{fig:deep_b}
    \end{subfigure}
    \begin{subfigure}{0.0\textwidth}
        \phantomcaption
        \label{fig:rome_c}
    \end{subfigure}
    \begin{subfigure}{0.0\textwidth}
        \phantomcaption
        \label{fig:rome_d}
    \end{subfigure}
\vspace{-0.5cm}
\caption{(a) deep triangulations (\texttt{orig}), (b) deep triangulations weighted by $H_\textrm{min}$, (c) Rome (\texttt{orig}), (d) Rome weighted by $H_\textrm{min}$}
\label{fig:deep_rome_results}
\vspace{-0.5cm}
\end{figure}

\paragraph*{Layouts}
We create layouts using the spring-based methods \emph{FA2} and \emph{SM}. For each grid or  triangulation $G$ and each spring-method $\mathcal S$, we compute seven layouts:

\begin{itemize}
\item $\texttt{orig} \equiv \mathcal{S}(G)$ -- spring-embedding of a graphs $G$,
\item $\texttt{on\_top}$ -- drawing $\mathcal{S}(G)$ with edges of $\overline{G}\setminus G$ appended on top of the drawing, where $\overline{G}$ is the augmented version of graphs $G$, 
\item $\texttt{redraw} \equiv \mathcal{S}(\overline{G})$ -- spring-embedding of $\overline{G}$, 
\item $H_\textrm{min} \equiv \mathcal{S}(\overline{G}_{\textrm{min}})$, $H_\textrm{max} \equiv \mathcal{S}(\overline{G}_{\textrm{max}})$, $H_\textrm{mean} \equiv \mathcal{S}(\overline{G}_{\textrm{mean}})$ -- spring-embeddings of $\overline{G}$ where outlier edges are weighted according to the functions min, max and mean, respectively (see Eqn.~\ref{eq:fv}), jointly referred to as $\texttt{heuristic}$ layouts,
\item $H_\textrm{nb} \equiv \mathcal{S}(\overline{G}_{\textrm{nb}})$ -- spring-embeddings of $\overline{G}$ where node-pairs are weighted using the neighborhood weighting function. 
\end{itemize}

For the deep triangulations and Rome graphs, which are challenging by themselves, we do not consider augmented versions, thus $\texttt{on\_top}$ and $\texttt{redraw}$ are not computed for them. We run FA2 and SM five times for each graph, and choose the best layout, w.r.t. to the number of crossings. Both SM and FA2 are run for a maximum of 2000 iterations with default settings.  

\paragraph*{Quality Metrics}
We quantitatively evaluate the results by computing three quality metrics: the crossing number, $\mathtt{nc}\in [0,\infty]$ -- the total number of crossings in a layout; the angular resolution, $\mathtt{ang\_res}\in [0,1]$ -- the minimum angle between any two incident edges normalized by $2\pi/\max_{v\in V}{\textrm{deg}(v)}$, and crossing resolution, $\mathtt{cros\_res}\in [0,1]$ -- the minimum angle of any two crossing edges normalized by $\pi / 2$. Additionally, in order to measure how the augmenting edges distort the layouts, we compute the Procrustes Statistic~\cite{procrustes}, $\mathtt{ps}\in [0,1]$. Here, a value of $0$ indicates that two layouts are exactly similar in the positions of vertices, after rotation, translation and scaling. Finally, we note that while stress\cite{stress_eval} is a common function to evaluate layout quality, we do not use it in our evaluation, since layouts that show well planar substructures in planar graphs do not have low stress, since in such layouts some edges are (much) longer than the optimal value dictated by stress. 


\section{Results and Discussion} \label{sec:results}

\begin{figure*}[h!]
   \includegraphics[width=1.0\linewidth]{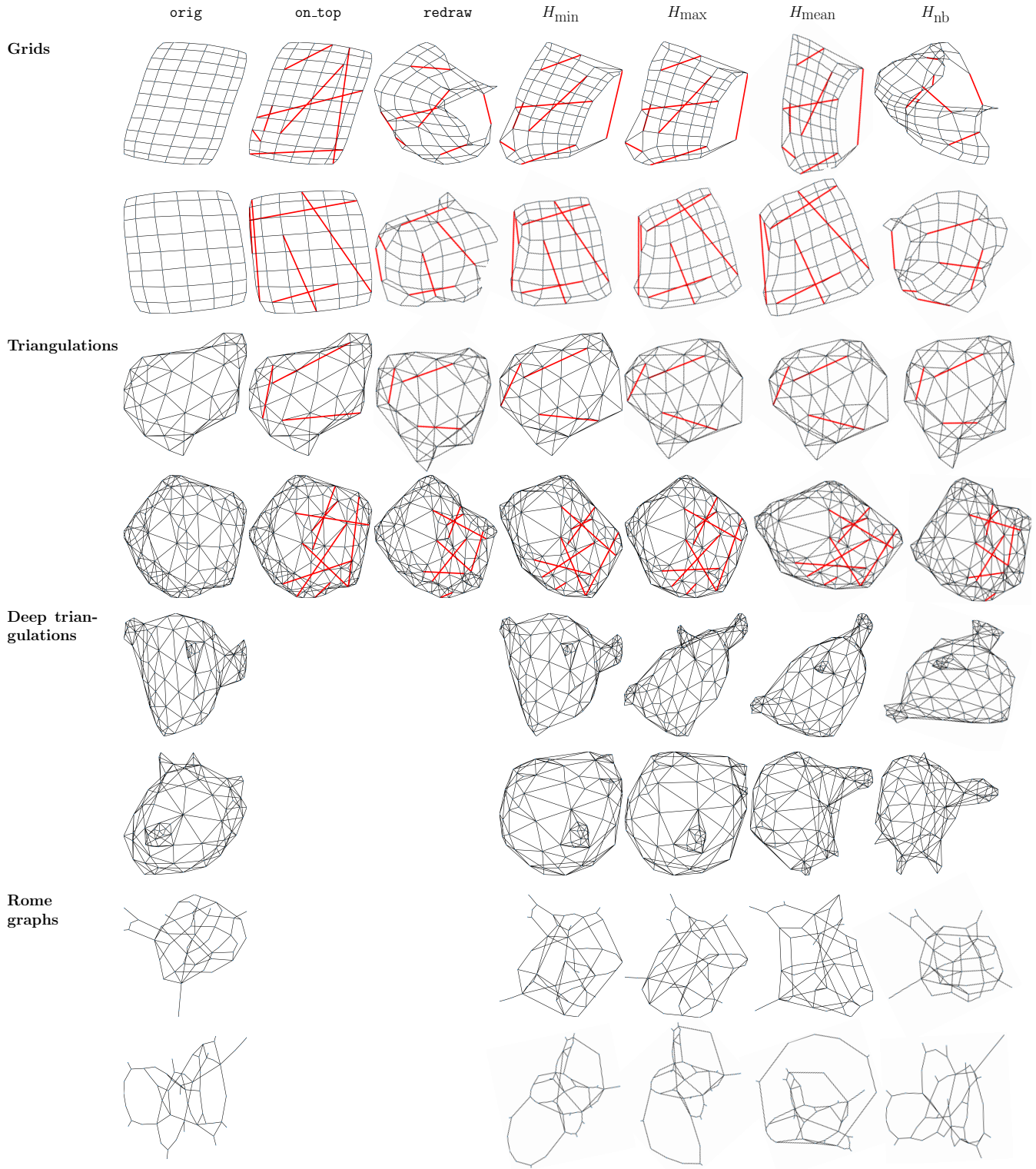}
        \caption{ForceAtlas2 layouts (left to right) \texttt{orig}, \texttt{on\_top}, \texttt{redraw} $H_\textrm{min}$, $H_\textrm{max}$, $H_\textrm{mean}$ and $H_\textrm{nb}$ for several graphs}
        \label{fig:full_page_image}
\end{figure*}

\paragraph*{Qualitative analysis}
Representative examples of running FA2 on augmented graphs (\texttt{redraw}) are depicted in Figures~\ref{fig:teaser_grid1}, \ref{fig:teaser_del1} and \ref{fig:full_page_image}. Here, the outer faces appear cluttered and the layouts appear to be folded inwards. Figures~\ref{fig:teaser_grid2} and \ref{fig:teaser_del2} show the layouts where edges are weighted according to the $\texttt{heuristic}$. We observe that the augmenting edges look longer, this makes the layouts being less folded inwards,  which in turn removes clutter and brings the grid-like and triangulation structures forward. 
The results of the \texttt{heuristic} on deep triangulations (Figure~\ref{fig:deep_a},~\ref{fig:deep_b}) also show some improvements in the clutter, especially in the outer face. Finally, for the Rome graphs (Figures~\ref{fig:rome_c}, \ref{fig:rome_d}), we do not see any improvement in the quality of the layout. However, this result is expected, as Rome graphs do not contain dense plain substructures.

\paragraph*{Quantitative analysis}

\begin{figure}
    \centering
    \includegraphics[width=1\linewidth]{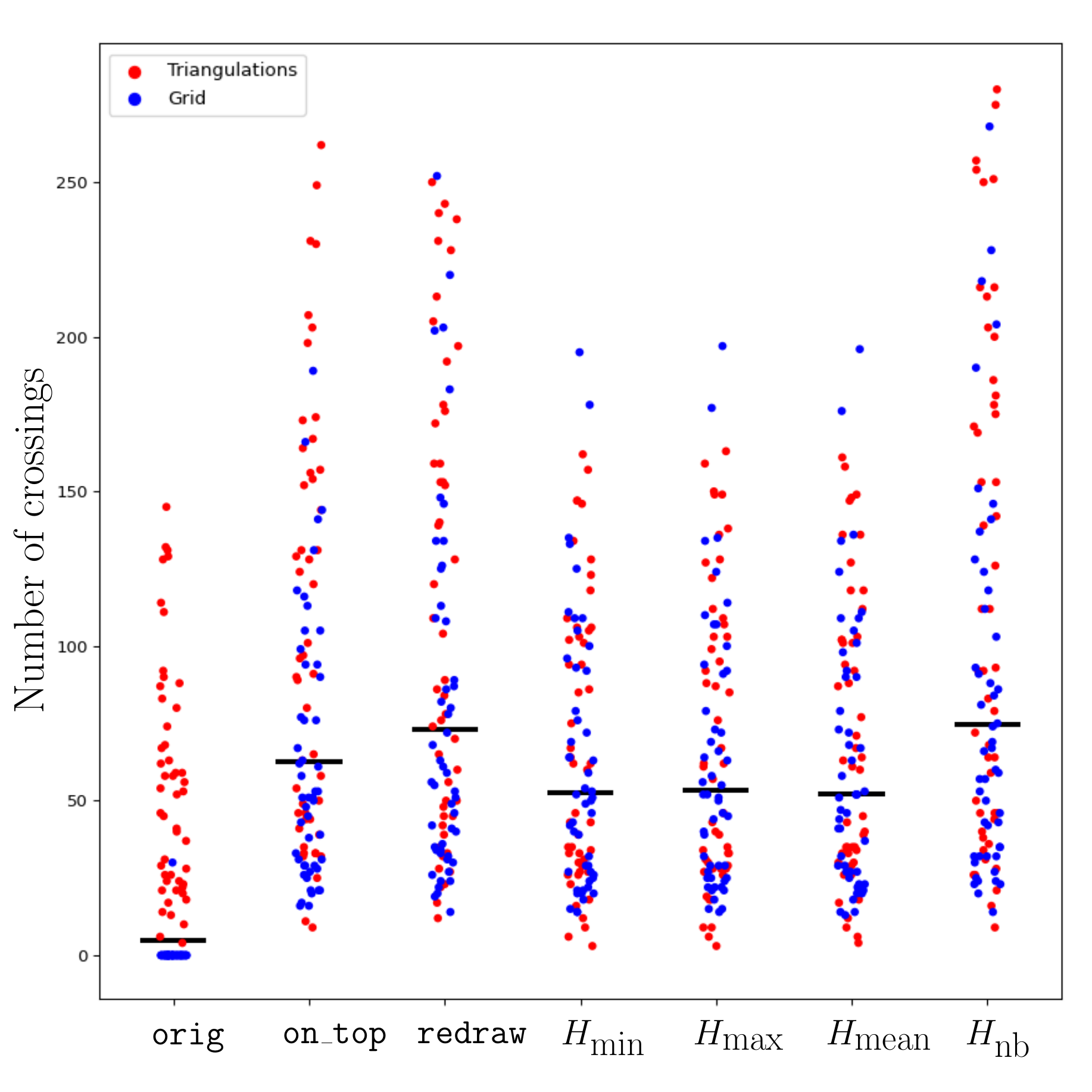}
    \caption{Number of crossings of grids and triangulations in FA2 layouts described in Section~\ref{sec:setup}. Black lines represent the median}
    \label{fig:jitter_grids_del}
\end{figure}

Figure~\ref{fig:jitter_grids_del} shows the distributions of the number of crossings of the seven layouts described in Section~\ref{sec:setup} for the augmented triangulations and grids (red and blue, respectively). We see that number of crossings in the \texttt{heuristic} layouts tend to be lower compared to the \texttt{redraw} layouts produced by pure spring-embedder. We also notice that the \texttt{nb} layouts, produced using neighborhood weighting function, show no change in the number of crossings, in comparison to the \texttt{redraw} layouts. 

Tables~\ref{tab:full_table} shows the median values of the quality metrics of layouts of all datasets. Since the data are paired but non-normally distributed, we use the two-sided Wilcoxon signed rank test to indicate significant ($\alpha = 0.05)$ improvement or deterioration in a quality metric. For the grids and triangulations we measure whether there is a significant difference between the \texttt{heuristic} and the \texttt{redraw} layouts. Whereas, for deep triangulations and Rome graphs we compare the \texttt{heuristic} with the \texttt{orig} layout.

Regarding the number of crossings, we observe significant improvements on the \texttt{heuristic} layouts of the augmented grids and (deep) triangulations graphs. Additionally, the \texttt{heuristic} approach outshines the neighborhood technique for these graphs. As expected from the qualitative analysis, for the Rome graphs the \texttt{heuristic} layouts are either equally good ($H_{\textrm{max}}$) or significantly worse ($H_\textrm{mean}$, $H_\textrm{min}$) w.r.t the number of crossings. 

Regarding angular resolution, we observe significant improvements over the \texttt{redraw} layouts in the augmented grids when FA2 is used. Whereas SM scores significantly worse for both the triangulations and Rome graphs.
For the crossing resolution, we observe no significant differences for the \texttt{heuristic} over the \texttt{orig} layouts for the augmented grids, triangulations and deep triangulations. 
However, the \texttt{heuristic} layouts are significantly worse for the Rome graphs, as  expected from the structure of the Rome graphs and the intention of our heuristic. 

Additionally, we note that the Procrustes Statistic values for the \texttt{heuristic} layouts tend to be close to 0, for all but the Rome dataset. These results indicate that the weighting tactic of our heuristic yields layouts that stay close to the original planar structure. Lastly, we observe that FA2 scores better than SM on most metrics for all datasets.

\begin{table}[t!]
\centering
\small
\begin{tabular}{||l||l|l||l|l||l|l||l|l||}
\hline
 \multicolumn{1}{||l}{\textbf{Grids}}  & \multicolumn{2}{c}{$\mathtt{nc}$} & \multicolumn{2}{c}{$\mathtt{ang\_res}$} & \multicolumn{2}{c}{$\mathtt{cros\_res}$} & \multicolumn{2}{c||}{$\mathtt{ps}$}\\
\hline
 $\texttt{orig}$ & \FAC 0 & \SMC 0 & \FAC .631 & \SMC .982 & \FAC - & \SMC - & \FAC - & \SMC -\\
 $\texttt{on\_top}$ & \FAC 52 & \SMC 52 & \FAC .011 & \SMC .005 & \FAC .10 & \SMC .16 & \FAC 0 & \SMC 0\\
 $\texttt{redraw}$ & \FAC 67 & \SMC 102 & \FAC .014 & \SMC .018 & \FAC .14 & \SMC .12 & \FAC .11 & \SMC .25\\
 $H_\textrm{mean}$ & \FACS 51 & \SMCS 52 & \FACS .027 & \SMC .020 & \FAC .14 & \SMC .14 & \FAC .02 & \SMC .02\\
 $H_\textrm{min}$ & \FACS 51 & \SMCS 52 & \FACS .030 & \SMC .020 & \FAC .11 & \SMC .14 & \FAC .02 & \SMC .02\\
 $H_\textrm{max}$ & \FACS 51 & \SMCS 52 & \FAC .028 & \SMC .020 & \FAC .14 & \SMC .14 & \FAC .02 & \SMC .02\\
 $H_\textrm{nb}$ & \FACS 60 & \SMCS 94 & \FAC .019 & \SMCS .026 & \FAC .17 & \SMC .12 & \FAC .11 & \SMC .19\\
\hline
 \multicolumn{9}{||l||}{\textbf{Triangulations}}\\
\hline
 $\texttt{orig}$ & \FAC 53 & \SMC 79 & \FAC .009 & \SMC .023 & \FAC .10 & \SMC .14 & \FAC - & \SMC -\\
 $\texttt{on\_top}$ & \FAC 96 & \SMC 131 & \FAC .008 & \SMC .016 & \FAC .08 & \SMC .10 & \FAC 0 & \SMC 0\\
 $\texttt{redraw}$ & \FAC 87 & \SMC 154 & \FAC .009 & \SMC .019 & \FAC .12 & \SMC .12 & \FAC .03 & \SMC .08\\
 $H_\textrm{mean}$ & \FACS 63 & \SMCS 88 & \FAC .015 & \SMCS .012 & \FAC .12 & \SMC .12 & \FAC .02 & \SMC .06\\
 $H_\textrm{min}$ & \FACS 63 & \SMCS 88 & \FAC .012 & \SMCS .012 & \FAC .11 & \SMC .12 & \FAC .02 & \SMC .06\\
 $H_\textrm{max}$ & \FACS 65 & \SMCS 88 & \FAC .013 & \SMCS .012 & \FAC .12 & \SMC .12 & \FAC .02 & \SMC .06\\
 $H_\textrm{nb}$ & \FACS 93 & \SMCS 124 & \FAC .013 & \SMC .015 & \FAC .10 & \SMC .09 & \FAC .04 & \SMC .07\\
\hline
 \multicolumn{9}{||l||}{\textbf{Deep triangulations}}\\
\hline
 $\texttt{orig}$ & \FAC 74 & \SMC 102 & \FAC .017 & \SMC .020 & \FAC .13 & \SMC .15 & \FAC - & \SMC -\\
 $H_\textrm{mean}$ & \FACS 59 & \SMCS 82 & \FAC .010 & \SMC .022 & \FAC .11 & \SMC .13 & \FAC .03 & \SMC .09\\
 $H_\textrm{min}$ & \FACS 61 & \SMCS 82 & \FAC .019 & \SMC .022 & \FAC .12 & \SMC .13 & \FAC .03 & \SMC .09\\
 $H_\textrm{max}$ & \FACS 58 & \SMCS 82 & \FAC .016 & \SMC .022 & \FAC .16 & \SMC .13 & \FAC .03 & \SMC .09\\
 $H_\textrm{nb}$ & \FAC 73 & \SMCS 91 & \FAC .017 & \SMC .023 & \FAC .10 & \SMC .13 & \FAC .03 & \SMC .04\\
\hline
 \multicolumn{9}{||l||}{\textbf{Rome}}\\
\hline
 $\texttt{orig}$ & \FAC 26 & \SMC 30 & \FAC .048 & \SMC .065 & \FAC .30 & \SMC .26 & \FAC - & \SMC -\\
 $H_\textrm{mean}$ & \FACS 28 & \SMCS 34 & \FAC .035 & \SMCS .029 & \FACS .21 & \SMCS .15 & \FAC .38 & \SMC .53\\
 $H_\textrm{min}$ & \FACS 29 & \SMCS 34 & \FAC .033 & \SMCS .029 & \FACS .18 & \SMCS .15 & \FAC .34 & \SMC .53\\
 $H_\textrm{max}$ & \FAC 24 & \SMCS 34 & \FAC .037 & \SMCS .029 & \FACS .21 & \SMCS .15 & \FAC .35 & \SMC .53\\
 $H_\textrm{nb}$ & \FAC 25 & \SMCS 25 & \FAC .052 & \SMC .062 & \FAC .27 & \SMC .29 & \FAC .13 & \SMC .18\\
\hline
\end{tabular}
\caption{Median values of metrics. For grids and triangulations, $\texttt{redraw}$ is compared with \texttt{heuristic} layouts. For deep triangulations and Rome, $\texttt{orig}$ is compared with \texttt{heuristic} layouts. A stronger hue indicates a significant result, with \FA{FA2} \& \SM{SM}
\label{tab:full_table}
}
\end{table}


\section{Conclusion}
We presented a heuristic to detect edges that create clutter in layouts of near planar graphs. By suitably weighting such edges, we use spring-embedders to draw these graphs with the goal to better convey their planar substructures. The experiments indicate that our heuristic produces better results for augmented grids and triangulations. For deep triangulations we noticed visual improvements and clutter decrease in the outer face, though further improvements are possible. Moreover, our heuristic produces drawings with fewer number of crossings than conventional methods for all but the Rome graphs. This result is, however, expected since the Rome graphs do not contain dense planar substructures. Future work can yield more insight into deep triangulations, which we expect to be very challenging to lay out in a way that reveal their planar structure. Moreover, additional comparisons can be made between our heuristic and tsNET$^\star$. In addition to more experiments, future work can attempt to improve the heuristic's limiting time complexity, by altering or substituting the vertex-disjoint path and outlier detection computations. Finally, we plan to test whether Graph Neural Networks can be more successful in identifying \prob 
 edges.

\newpage
\bibliographystyle{eg-alpha-doi}
\bibliography{outlier_detection}



\end{document}